\documentclass[%
12pt]{article}
\usepackage{amssymb}
\usepackage{amsmath, epsfig, rotating}

\newcommand{\be}{\begin{equation}}
\newcommand{\en}{\end{equation}}

\renewcommand{\vec}[1]{\boldsymbol{#1}}

\newcommand{\demi}{\textstyle{\frac{1}{2}}}

\begin{document}

\title{On the third- and fourth-order constants \\
of incompressible isotropic elasticity}


\author{%
Michel Destrade$^a$, Raymond W. Ogden$^{b}$\\[12pt]
%
$^a$School of Electrical, Electronic, and Mechanical Engineering,\\
University College Dublin,\\ Belfield, Dublin 4, Ireland.\\[12pt]
$^b$Department of Mathematics,\\
University of Glasgow,\\
University Gardens,
Glasgow G12 8QW,
Scotland, UK.}

\date{}

\maketitle

\begin{abstract}
Consider the constitutive law for an isotropic elastic solid with the strain-energy function expanded up to the fourth order in the strain, and the stress up to the third order in the strain.
The stress-strain relation can then be inverted to give the strain in terms of the stress with a view to considering the incompressible limit.
For this purpose, use of the logarithmic strain tensor is of particular value.
It enables the limiting values of all nine fourth-order elastic constants in the incompressible limit to be evaluated precisely and rigorously.
In particular, it is explained why the three constants of fourth-order incompressible elasticity $\mu$, $\bar{A}$, and $\bar D$ are of the same order of magnitude.
Several examples of application of the results follow, including determination of the acoustoelastic coefficients in incompressible solids and the limiting values of the coefficients of nonlinearity for elastic wave propagation.

\end{abstract}
\newpage




\section{INTRODUCTION}


Extracting the condition for incompressibility from a stress-strain relation can be an ambiguous process because it leads to an \emph{infinite} limit for one (or more) of the elastic stiffnesses and eventually, to the appearance of a hydrostatic stress term proportional to an \emph{arbitrary} Lagrange multiplier (to be determined from boundary and/or initial conditions)\cite{Dest01,OgVi04}.
In linear isotropic elasticity, the relation between the infinitesimal stress $\vec{\sigma}$ and the infinitesimal strain $\vec{\epsilon}$ reads
\be
\vec{\sigma} = \lambda\, \text{tr} (\vec{\epsilon})\vec{\delta} + 2\mu  \vec{\epsilon},
\en
where $\lambda$ and $\mu$ are the Lam\'e coefficients and $\vec{\delta}$ is the identity.
The incompressible limit is equivalent to the condition $\text{tr}\ \vec{\epsilon} = 0$, which leads to the limiting case
\be
\lambda \rightarrow \infty, \qquad
\vec{\sigma} = -p \vec{\delta} + 2\mu \vec{\epsilon},
\en
where $p$ is an arbitrary scalar.
By way of contrast, a strain--stress relation is more amenable to the imposition of incompressibility, because it leads to unambiguous, \emph{finite} limit(s) for one (or several) compliance(s)\cite{DeMT02}.
Hence, in linear isotropic elasticity, we go from
\be
\vec{\epsilon} = -\dfrac{\nu}{E}\,\text{tr}(\vec{\sigma})\vec{\delta} + \dfrac{1+\nu}{E}\vec{\sigma}, \label{strain-stress-linear}
\en
where $\nu$ is Poisson's ratio and $E$ is Young's modulus, to the limiting case
\be
\nu \rightarrow \dfrac{1}{2}, \qquad
\vec{\epsilon} = -\dfrac{1}{2E}\,\text{tr}(\vec{\sigma})\vec{\delta} + \dfrac{3}{2E}\vec{\sigma},
\en
and there is no arbitrary quantity.

Turning now to \emph{nonlinear} isotropic elasticity, we must first of all make a choice of the measures of stress and of strain.
Physicists and acousticians seem to favor the pair consisting of the Green--Lagrange strain tensor $\vec{\bar{e}}$ and the second Piola--Kirchhoff stress $\vec{\bar{t}}$, and they expand, in the so-called weakly nonlinear theory,  the strain-energy density $W$ in terms of three isotropic invariants of the strain.
Hence, the Landau and Lifshitz\cite{LaLi86} expansion can be conducted to fourth-order as\cite{Zabo86}
\begin{equation} \label{fourth}
W =
\dfrac{\lambda}{2}\bar{I}_1^2 +\mu \bar{I}_2 + \dfrac{\bar{A}}{3} \bar{I}_3 + \bar{B} \bar{I}_1 \bar{I}_2
+  \dfrac{\bar{C}}{3}\bar{I}_1^3
+ \bar{E}  \bar{I}_1\bar{I}_3
+ \bar{F}  \bar{I}_1^2 \bar{I}_2
+ \bar{G}  \bar{I}_2^2
+ \bar{H} \bar{I}_1^4,
\end{equation}
where $\bar{A}$, $\bar{B}$, $\bar{C}$ are the three third-order constants, $\bar{E}$, $\bar{F}$, $\bar{G}$, $\bar{H}$ are the four fourth-order constants, and
\be \label{I_k}
\bar{I}_1 = \text{tr} (\vec{\bar{e}}), \qquad
\bar{I}_2 = \text{tr}(\vec{\bar{e}}^2), \qquad
\bar{I}_3 = \text{tr} (\vec{\bar{e}}^3),
\en
are respectively of first, second, and third order in the strain. (Note that we have placed overbars on the elastic constants so as to avoid conflict with the standard notation $E$ for Young's modulus, which is used extensively in what follows.)
Thus, terms in $\bar{I}_1^2$ and $\bar{I}_2$ are ``second-order'' terms; terms in $\bar{I}_1^3$, $\bar{I}_1 \bar{I}_2$, $\bar{I}_3$ are ``third-order'' terms, and so on.
However, the incompressibility limit is not easily implemented with this choice, because it is a combination of the invariants\cite{JCGB07}
, specifically
\be
\bar{I}_1 - \bar{I}_2 + \bar{I}_1^2 + \tfrac{2}{3} \bar{I}_1^3 - 2 \bar{I}_1 \bar{I}_2 + \tfrac{4}{3}\bar{I}_3 = 0.
\en
This, in particular, means that $\bar{I}_1$ is now a second-order quantity.  This makes it complicated to arrive at the fourth-order expansion of incompressible nonlinear elasticity, for which the strain-energy function has the form
\begin{equation} \label{fourth-incomp}
W = \mu \bar{I}_2 + \dfrac{\bar{A}}{3} \bar{I}_3 +  \bar{D} \bar{I}_2^2,
\end{equation}
where $\bar{D}$ is another constant.

However, it is advantageous to use, instead of Green--Lagrange strain, an alternative measure of strain, namely the \emph{logarithmic strain}, since this leads to a very simple expression of the incompressibility constraint. The logarithmic strain $\vec{e}$ is defined by
\be \label{log}
\vec{e} = \demi \ln(\vec{\delta} + 2\vec{\bar{e}}),
\en
and we consider the three independent invariants
\be
I_1 = \text{tr} ( \vec{e}), \qquad
I_2 = \text{tr}(\vec{e}^2), \qquad
I_3 = \text{tr} (\vec{e}^3),
\en
respectively of orders one, two, and three,
analogously to $\bar{I}_1$, $\bar{I}_2$, $\bar{I}_3$.
Then incompressibility is expressed \emph{exactly} as
\be \label{exact}
I_1=0,
\en
which must hold identically for all stresses and strains.

These considerations suggest the following protocol for finding the limiting values of the elastic constants in incompressible weakly nonlinear elasticity (Sections \ref{section3} and \ref{section4}).
First, write down the relation between stress and strain.  
Since we are considering isotropic materials, the relevant stress measure $\vec{t}$ conjugate to logarithmic strain $\vec{e}$ is the Kirchhoff stress tensor on rotated coordinates, related to $\vec{\bar{t}}$ by $\vec{t} = (\vec{\delta} + 2 \vec{\bar{e}})\vec{\bar{t}}$ (for a general discussion of conjugate stress and strain tensors see Ogden\cite{Ogde84}).
Then invert this relation to find the strain in terms of the stress, and, more particularly, $I_1$ in terms of invariants of $\vec{t}$. Finally, note that $I_1$ must be zero for all $\vec{t}$.
In what follows we conduct this process explicitly for third-order elasticity.
Then we present the results of the fourth-order case, omitting many of the (cumbersome and lengthy) details of the calculations.
In Section \ref{examples}, we give a few examples of applications, and we see that results established in compressible elasticity can be taken to their incompressible limit, without having to re-write and to re-solve the equations of motion and the boundary conditions.

Previously, Hamilton \emph{et al.}\cite{HaIZ04} have shown that there should remain only three elastic constants in the incompressible limit of Eq. \eqref{fourth} (Ogden\cite{Ogde74}  had in fact proved this result 30 years earlier).
They also found some limited information on the behavior of the other six constants, by extending the work of Kostek \emph{et al.}\cite{KoSN93} from third-order to fourth-order elasticity.
Based on a comparison with the equation of state of inviscid fluids, it was found that $\mu$ and $\bar{A}$ remain, a new fourth-order constant $\bar{D}$ emerges, and
\begin{align} \label{set}
&\lambda \rightarrow \infty, \qquad \bar{B} = - \lambda, \qquad
\bar{E} = \frac{4 \lambda}{3}, \notag \\ & \bar{F} = -\bar{C}, \qquad
\bar{G}  = \frac{\lambda}{2}, \qquad \bar{D} = \dfrac{\lambda}{2} + \bar{B} + \bar{G};
\end{align}
see Zabolotskaya \emph{et al.}\cite{HaIZ04B} and Jacob \emph{et al.}\cite{JCGB07} for the latter identity.
However, the behavior of $\bar{C}$, $\bar F$, and $\bar{H}$ remains undetermined.
Moreover, the full comparison with fluids ultimately leads to the identities: $\mu=0$, $\bar{A}=0$, $\bar{D}=0$, which, in considering an elastic solid, are not satisfactory because they imply that some information might be missing from the list \eqref{set}, where $\mu$, $\bar{A}$, and $\bar{D}$ could play a role.
One can only go so far in comparing the behavior of a solid to that of a fluid:
a particular disconnect emerges when comparing the behavior of a solid in the incompressible limit, where the speed of a longitudinal wave should tend to infinity, to that of an isentropic liquid, where the speed of a longitudinal wave is finite (Section \ref{Nonlinear-plane-waves}).
The differences existing between a compressible solid and its incompressible counterpart must be tackled within the framework of solid mechanics.

In effect, many questions have remained open in the literature, as attested by the following comments.
Doma\'nski\cite{Doma09} remarks that ``the details of the derivation [of Eq. \eqref{fourth-incomp}] are not quite clear from the mathematical point of view'', and that ``surprisingly, the experiments confirm that, in spite of being a combination of the higher order constants, the fourth-order constant $\bar{D}$ is of a similar order of magnitude as the second-order shear Lam\'e constant $\mu$ and the third-order Landau constant $\bar{A}$''.
Catheline \emph{et al.}\cite{CaGF03} measure $\bar{A}$ and $\bar{B}$ for Agar-gelatin based phantoms and note that ``the huge difference between these third-order moduli is striking since in more conventional media such as metal, rocks, or crystals they are of the same order.'' They then propose an ``intuitive justification'' for this difference, which ``does not hold for a
theoretical explanation''.
Jacob \emph{et al.}\cite{JCGB07} record that ``surprisingly, the experiments confirm that, although expressed as function of compression moduli $\lambda$, $\bar{B}$, $\bar{G}$, the moduli $\bar{D}$ is of the order of magnitude of the so-called shear moduli $\bar{\mu}$ and $\bar{A}$,'' and add that ``no explanation has been given for this order of magnitude.''
We address each of these points in the course of this paper.


\section{THIRD-ORDER INCOMPRESSIBILITY}
\label{section3}


In this section we focus on third-order elasticity, so that the strain-energy function \eqref{fourth} reduces to
\begin{equation} \label{third}
W =
\dfrac{\lambda}{2}\bar{I}_1^2 +\mu \bar{I}_2 + \dfrac{\bar{A}}{3} \bar{I}_3 + \bar{B} \bar{I}_1 \bar{I}_2
+  \dfrac{\bar{C}}{3}\bar{I}_1^3.
\end{equation}
At this order, the inversion of Eq. \eqref{log} gives $\vec{\bar{e}} = \vec{e} + \vec{e}^2 + 2\vec{e}^3/3$, so that
\be \label{I-Ibar}
\bar{I}_1 = I_1 + I_2 + \tfrac{2}{3}I_3, \qquad
\bar{I}_2 = I_2 + 2 I_3, \qquad
\bar{I}_3 = I_3.
\en
It follows that the third-order expansion of $W$ in terms of the invariants of the logarithmic strain $\vec{e}$ reads
\begin{equation} \label{third-ea}
W =
\dfrac{\lambda}{2} I_1^2+ \mu I_2 + \left(\dfrac{\bar{A}}{3} + 2\mu \right) I_3 + (\bar{B}+\lambda) I_1 I_2 + \dfrac{\bar{C}}{3} I_1^3,
\end{equation}
which is also written as
\begin{equation} \label{third-e}
W =
\dfrac{E \nu}{2(1+\nu)(1-2\nu)} I_1^2 + \dfrac{E}{2(1+\nu)} I_2 + \dfrac{\mathcal{A}}{3} I_3 + \mathcal{B} I_1 I_2 + \dfrac{\mathcal{C}}{3} I_1^3,
\end{equation}
where $E$, $\nu$ are second-order constants (Young's modulus and Poisson's ratio) and $\mathcal{A}$, $\mathcal{B}$, $\mathcal{C}$ are third-order constants with respect to the logarithmic strain, with the connections
\begin{align} \label{A-Acal}
& E = \dfrac{3 \lambda + 2 \mu}{\lambda+\mu} \mu, \qquad
\nu = \dfrac{\lambda}{2(\lambda+\mu)}, \notag \\
& \mathcal{A} = \bar{A} + 6 \mu, \qquad
\mathcal{B} = \bar{B} + \lambda, \qquad
\mathcal{C} = \bar{C}
\end{align}
to the Lam\'e and Landau coefficients.
The (conjugate) stress $\vec{t} = \partial W/\partial \vec{e}$ then expands as
\be \label{t3}
\vec{t} = \left(\dfrac{E \nu}{(1+\nu)(1-2\nu)}I_1 + \mathcal{B}I_2 + \mathcal{C}I_1^2 \right)\vec{\delta}
 + \left( \dfrac{E}{1+\nu} + 2\mathcal{B}I_1\right)\vec{e} + \mathcal{A} \vec{e}^2.
 \en

Now we introduce the stress invariants $T_1 = \text{tr}(\vec{t})$ and $T_2 = \text{tr}(\vec{t}^2)$,
and it follows by taking in turn the trace of Eq. \eqref{t3} and then of its square that
\begin{align}
& T_1 = \dfrac{E}{1-2\nu} I_1 + (2\mathcal{B}+3\mathcal{C})I_1^2 + (\mathcal{A}+3\mathcal{B})I_2, \notag \\
& T_2 = \dfrac{E^2}{(1+\nu)^2}\left[\dfrac{\nu(2-\nu)}{(1-2\nu)^2} I_1^2 + I_2 \right], \notag \\
& T_1^2 = \dfrac{E^2}{(1-2\nu)^2} I_1^2,\label{stress-invars2}
\end{align}
correct to the second order in the strain.
These can be inverted, to give $I_1^2$, $I_2$, and then
\be\label{third-order_expansion}
I_1 = \alpha T_1 + \beta T_2 + \gamma T_1^2,
\en
where the constants $\alpha,\beta,\gamma$ are given by
\begin{align} \label{alpha}
& \alpha = \dfrac{1-2\nu}{E}, \notag \\
& \beta = -(\mathcal{A}+3\mathcal{B})\dfrac{(1+\nu)^2(1-2\nu)}{E^3}, \notag \\
& \gamma = (\mathcal{A}+3\mathcal{B})\dfrac{\nu(2-\nu)(1-2\nu)}{E^3} - (2\mathcal{B}+3\mathcal{C})\dfrac{(1-2\nu)^3}{E^3}.
\end{align}

For the exact incompressibility condition Eq. \eqref{exact} to hold for all stresses and strains, we must have $\alpha = \beta = \gamma =0$.
First, $\alpha = 0$ is clearly equivalent to either of
\be \label{lame-limit}
\nu \rightarrow 1/2, \qquad \lambda \rightarrow \infty,
\en
whilst both $E$ and $\mu$ remain finite, with $\mu \rightarrow E/3$.

Second, note that the stress must remain finite in the incompressible limit.
Hence, the last term in the expression for $\vec{t}$ in Eq. \eqref{t3} remains finite, so that $\mathcal{A}$ remains finite:
\be \label{lim-Acal}
\mathcal{A}/\mu = \mathcal{O}(1).
\en

With these two conditions for incompressibility, the expression for $\beta$ in Eq. \eqref{alpha} reduces to
\be
 \beta = - \dfrac{27\mathcal{B}(1-2\nu)}{4E^3}. \notag
\en
Notice that the first term in the expression for $\vec{t}$ in Eq. \eqref{t3} must also remain finite, which means that $(1-2\nu)^{-1}I_1$ remains finite.
Clearly, the conditions ``$\beta=0$'' and ``$(1-2\nu)^{-1}\beta$ is finite'' are fulfilled simultaneously when $\mathcal{B}$ remains finite:
\be \label{lim-Bcal}
\mathcal{B}/\mu = \mathcal{O}(1).
\en

With the three conditions for incompressibility Eqs. \eqref{lame-limit}--\eqref{lim-Bcal}, the expression for $\gamma$ in Eq. \eqref{alpha} reduces to
\be
 \gamma = - \dfrac{3\mathcal{C}(1-2\nu)^3}{E^3}.
\en
Recall that $(1-2\nu)^{-1}I_1$ and hence, $(1-2\nu)^{-1}\gamma$ remain finite.
It follows that
\be \label{lim-Ccal}
(1-2\nu)^3 \mathcal{C} \rightarrow 0, \qquad \mathcal{C}/\mu=\mathcal{O}(\lambda^2/\mu^2).
\en
In summary, the conditions given in Eqs. \eqref{lame-limit}, \eqref{lim-Acal},  \eqref{lim-Bcal}, and  \eqref{lim-Ccal} are necessary and sufficient for incompressibility.

For an incompressible material (with $I_1 \equiv 0$) the strain energy $W$ therefore reduces to
\be
W = \frac{E}{3} I_2 + \frac{\mathcal{A}}{3} I_3,
\en
or equivalently, in terms of the invariants of the Green strain tensor,
\be \label{incomp_W}
W = \frac{E}{3} (\bar{I}_2 - 2\bar{I_3}) + \frac{\mathcal{A}}{3} \bar{I}_3 = \mu \bar{I}_2 + \frac{\bar{A}}{3}\bar{I}_3,
\en
where we have used Eqs. \eqref{I-Ibar}--\eqref{A-Acal}.
We use Eqs. \eqref{A-Acal} to find that, in terms of Lam\'e and Landau constants, the incompressible limits are Eqs. \eqref{lame-limit} together with
\begin{align} \label{incomp_3rd}
& \bar{A} = \mathcal{A} - 6\mu = \mu\mathcal{O}(1), \qquad
  \bar{B} = \mathcal{B} - \lambda = \lambda\mathcal{O}(1),\qquad
  \bar{C} = \mathcal{C} = \mu\mathcal{O}(\lambda^2/\mu^2),\notag \\
& (1-2\nu)\bar{B} \rightarrow -E/3, \qquad
  (1-2\nu)^3 \bar{C} \rightarrow 0.
\end{align}
Notice that $\bar C$ varies in a quadratic manner with respect to $\lambda$ as it goes to infinity in incompressible solids, \emph{not} in a linear manner, as is incorrectly reported in Refs.\cite{HaIZ04, HaIZ07, OSIR07}; see also Section \ref{Nonlinear-plane-waves}, where it is shown that if $\bar C$ were linear in $\lambda$ then longitudinal waves would propagate with finite speed in the incompressible limit.

Bearing in mind the advantage of the strain--stress relation \eqref{strain-stress-linear} in the linear theory we now take note of its counterpart for the present theory.  On use of the definitions of $T_1$ and $T_2$ and Eq. \eqref{stress-invars2} we may, after lengthy algebra, invert the stress-strain equation \eqref{t3} to give
\begin{multline}
\vec{e} =  \dfrac{1+\nu}{E}\vec{t}-\dfrac{\nu}{E}T_1\vec{\delta}-\frac{\mathcal{A}(1+\nu)^3}{E^3}\vec{t}^2
+ \dfrac{(1+\nu)^2}{E^3} \left[\nu \mathcal{A}-(1-2\nu)\mathcal{B} \right] \left(2T_1\vec{t}+T_2\vec{\delta}\right) \\
 + \dfrac{1}{E^3}\left[3\mathcal{B}\nu (2-\nu)(1-2\nu)-3\mathcal{A}\nu^2-\mathcal{C}(1-2\nu)^3\right]T_1^2\vec{\delta}.\label{e3}
\end{multline}
In the incompressible limit as embodied in Eqs. \eqref{lame-limit}--\eqref{lim-Ccal}, this reduces to
\begin{equation}
\vec{e}=\frac{3}{4E^3}\left(2E^2+3\mathcal{A}T_1\right)\left(\vec{t}-\tfrac{1}{3}T_1\vec{\delta}\right) - \frac{27\mathcal{A}}{8E^3}\left(\vec{t}^2-\tfrac{1}{3}T_2\vec{\delta}\right),\label{e3incomp}
\end{equation}
from which it follows immediately that $\mbox{tr}(\vec{e})=0$.

Conversely, the stress-strain relation must accommodate the internal constraint of incompressibility by introducing a Lagrange multiplier, denoted $p$, in the following expansions,
\be \label{3stress-strain}
\vec{t} = -p \vec{\delta} + \dfrac{2E}{3}\vec{e} + \mathcal{A} \vec{e}^2, \qquad
 \vec{\bar{t}} = - p(\vec{\delta}+2\vec{\bar{e}})^{-1} + 2 \mu \vec{\bar{e}} + \bar{A} \vec{\bar{e}}^2,
\en
for the stresses, where we recall that for an isotropic material, $\vec{t} = (\vec{\delta} + 2 \vec{\bar{e}})\vec{\bar{t}}$.


\section{FOURTH-ORDER INCOMPRESSIBILITY}
\label{section4}


We now extend the above analysis to include fourth-order terms in the strain-energy function, as in \eqref{fourth}, and we work in terms of the logarithmic strain tensor and its invariants.
We use the fourth-order expansion $\vec{\bar{e}}=\vec{e}+\vec{e}^2+2\vec{e}^3/3+\vec{e}^4/3$  and the identity\cite{HaIZ04}
$ \text{tr}(\vec{e}^4)= (1/6) I_1^4 - I_1^2I_2 + (1/2) I_2^2 + (4/3) I_1 I_3$ to establish the following connections between the invariants of $\vec{\bar{e}}$ and those of $\vec{e}$:
\begin{align}
& \bar{I}_1 = I_1 + I_2 + \tfrac{2}{3}I_3 + \tfrac{4}{9}I_1 I_3 - \tfrac{1}{3}I_1^2I_2 + \tfrac{1}{6}I_2^2 + \tfrac{1}{18}I_1^4, \notag \\
& \bar{I}_2 = I_2 + 2 I_3 - \tfrac{7}{3}I_1^2I_2 + \tfrac{28}{9}I_1 I_3 + \tfrac{7}{6}I_2^2 + \tfrac{7}{18}I_1^4, \notag \\
& \bar{I}_3 = I_3 + 4 I_1 I_3 - 3I_1^2 I_2 + \tfrac{3}{2}I_2^2 + \tfrac{1}{2}I_1^4.
\end{align}

The strain energy may then be expressed as a function of the invariants of $\vec{e}$, in the form
\begin{equation} \label{4thlog}
W =
\dfrac{\lambda}{2}{I}_1^2 +\mu {I}_2 + \dfrac{\mathcal{A}}{3} {I}_3 + \mathcal{B} {I}_1 {I}_2
+  \dfrac{\mathcal{C}}{3}{I}_1^3
+ \mathcal{E} {I}_1{I}_3
+ \mathcal{F}  {I}_1^2 {I}_2
+ \mathcal{G} {I}_2^2
+ \mathcal{H} {I}_1^4,
\end{equation}
where, in addition to the connections \eqref{A-Acal}, the elastic constants are given by
\begin{align}
&  \mathcal{E} = \bar{E} + 2\bar{B} + \dfrac{4 \bar{A}}{3} + \dfrac{28 \mu}{9} + \dfrac{2\lambda}{3},
&& \mathcal{F} = \bar{F} + \bar{C} - \bar{A} - \dfrac{7 \mu}{3},
\notag \\
&  \mathcal{G} = \bar{G} + \bar{B} + \dfrac{\lambda}{2} + \dfrac{\bar{A}}{2} + \dfrac{7\mu}{6},
&& \mathcal{H} = \bar{H} + \dfrac{\bar{A}}{6} + \dfrac{7\mu}{18}.
\end{align}
Note that it is a simple matter to invert these relations using Eqs. \eqref{A-Acal}, to give
\begin{align} \label{connec4}
&  \bar{E} = \mathcal{E} - 2\mathcal{B} - \dfrac{4 \bar{A}}{3} + \dfrac{44 \mu}{9} + \dfrac{4\lambda}{3},
&& \bar{F} = \mathcal{F} - \mathcal{C} + \mathcal{A} - \dfrac{11 \mu}{3},
\notag \\
&  \bar{G} = \mathcal{G} - \mathcal{B}  - \dfrac{\mathcal{A}}{2} + \dfrac{11 \mu}{6} + \dfrac{\lambda}{2},
&& \bar{H} = \mathcal{H} - \dfrac{\mathcal{A}}{6} + \dfrac{11 \mu}{18}.
\end{align}

The stress $\vec{t}$, to third order in $\vec{e}$, is then given by
\begin{multline}
\vec{t} = \lambda I_1\vec{\delta}+2\mu\vec{e} +\mathcal{B}(I_2\vec{\delta}+2I_1\vec{e})+\mathcal{A}\vec{e}^2+\mathcal{C}I_1^2\vec{\delta} \\
+ \mathcal{E}(I_3\vec{\delta} + 3I_1\vec{e}^2) + 2\mathcal{F}(I_1I_2\vec{\delta} + I_1^2\vec{e}) + 4\mathcal{G}I_2\vec{e} + 4\mathcal{H}I_1^3\vec{\delta}.
\label{stress4}
\end{multline}
By defining $T_1=\mbox{tr}(\vec{t})$, $T_2=\mbox{tr}(\vec{t}^2)$, $T_3=\mbox{tr}(\vec{t}^3)$, similarly to the previous section, by manipulating the equations to third order, and after some considerable algebra (which is omitted), we obtain an extension of the expansion \eqref{third-order_expansion} to give
\begin{equation}
I_1=\alpha T_1 + \beta T_2 + \gamma T_1^2 +\alpha' T_3 +\beta' T_1T_2 +\gamma' T_1^3,
\end{equation}
where $\alpha,\beta,\gamma$ are as given by \eqref{alpha}, while $\alpha',\beta',\gamma'$ are defined via
\begin{eqnarray}
E^5\alpha'&= & (1+\nu)^3(1-2\nu)\left[2(1+\nu)\mathcal{A}(\mathcal{A}+3\mathcal{B}) - 3E \mathcal{E}\right],
\notag\\
E^5\beta' &=  & (1 + \nu)^2 (1 - 2\nu) \Big\{2(\mathcal{A} + 3\mathcal{B}) \left[(1-2\nu)(5-4\nu)\mathcal{B}+3(1-2\nu)^2\mathcal{C} \right. \notag\\
&& \qquad\left. - \nu(4 + \nu)\mathcal{A}\right]
   - E\left[3(1  - 5\nu) \mathcal{E} + 2(1-2\nu)(3\mathcal{F} + 2\mathcal{G})\right] \Big\},
\notag \\
E^5\gamma' &= & 2\nu^2(2-\nu)(1-2\nu)^2(\mathcal{A}+3\mathcal{B})^2
\notag \\
&& -6\nu(1-\nu)(1-2\nu)^3(\mathcal{A}+3\mathcal{B})(2\mathcal{B}+3\mathcal{C})
\notag\\
&&  + \,2(1-2\nu)^5(2\mathcal{B} 
+ 3\mathcal{C})^2\notag \\
&&  + 3\nu^2(1-2\nu)[2(1+\nu)\mathcal{A}(\mathcal{A}+3\mathcal{B})-3 E\mathcal{E}]
\notag\\
&& - \, \nu(2-\nu)(1-2\nu)^2[6(1+\nu)(\mathcal{A}+3\mathcal{B})\mathcal{B} - E(3\mathcal{E}+6\mathcal{F}+4\mathcal{G})]
\notag\\
&& + \, 2(1-2\nu)^4[(1+\nu)(\mathcal{A}+3\mathcal{B})\mathcal{C} - E(\mathcal{F}+6\mathcal{H})].
\end{eqnarray}

The exact incompressibility constraint Eq. \eqref{exact} is enforced for all stresses when $\alpha = \beta = \gamma = 0$ and $\alpha' = \beta' = \gamma' = 0$.
The first three conditions lead to the limits Eqs. \eqref{lame-limit}--\eqref{lim-Ccal}, thereby bringing great simplifications in the expressions above.
Hence in the incompressible limit we have
\be
E^4 \alpha' = - \tfrac{81}{8}(1-2\nu)\mathcal{E},
\en
which must tend to zero as $\nu \rightarrow 1/2$.
However, on inspection of \eqref{stress4} it can be seen immediately that for the stress to remain finite in the incompressible limit we must have that $\mathcal{E}$ is finite:
\be \label{lim-Ecal}
\mathcal{E} / \mu=\mathcal{O}(1),
\en
and $\alpha'$ above does indeed tend to zero as $\nu \rightarrow 1/2$.
As in the previous section, we see that the first term in Eq. \eqref{stress4} remains finite in the incompressible limit when $(1-2\nu)^{-1}I_1$ remains finite.
Clearly here, $(1-2\nu)^{-1} \alpha'$ remains finite.

Equally, for the term in $\mathcal{G}$ in Eq. \eqref{stress4} to remain finite, $\mathcal{G}$ must itself be finite in the limit:
\be \label{lim-Gcal}
\mathcal{G} / \mu = \mathcal{O}(1),
\en
Now we find that in the incompressible limit, $\beta'$ behaves as
\be
E^4 \beta' = - \tfrac{27}{2}(1-2\nu)^2\mathcal{F}.
\en
For $\beta'$ to tend to zero, and $(1-2\nu)^{-1}\beta'$ to remain finite, we must enforce the following behavior for $\mathcal{F}$:
\be \label{lim-Fcal}
(1-2\nu)^2 \mathcal{F} \rightarrow 0, \qquad \mathcal{F}/\mu=\mathcal{O}(\lambda/\mu).
\en
In these limits, the $I_1I_2\mathcal{F}$-term in the expression Eq. \eqref{stress4} for the stress remains finite, whilst the $I_1^2 \mathcal{F}$-term vanishes.

It remains to consider $\mathcal{H}$.
Using the limits above, we see that $\gamma'$ behaves as
\be
E^4 \gamma' = - 12(1-2\nu)^4\mathcal{H}.
\en
For this to tend to zero, and $(1-2\nu)^{-1}\gamma'$ to remain finite, we require
\be \label{lim-Hcal}
(1-2\nu)^4 \mathcal{H} \rightarrow 0, \qquad \mathcal{H}/\mu=\mathcal{O}(\lambda^3/\mu^3).
\en
We may then check that the limiting value of the last term in Eq. \eqref{stress4} is finite.

In summary, we must have for the fourth-order constants associated with the logarithmic strain
\begin{equation}
\mathcal{E}/\mu=\mathcal{O}(1),\qquad
\mathcal{F}/\mu=\mathcal{O}(\lambda/\mu),\qquad
\mathcal{G}/\mu=\mathcal{O}(1),\qquad
\mathcal{H}/\mu=\mathcal{O}(\lambda^3/\mu^3),
\end{equation}
which are necessary and sufficient for incompressibility at this order.

For the fourth-order constants associated with the Green strain, we use Eqs. \eqref{connec4} to find
\begin{equation}
\bar{E}/\mu=\mathcal{O}(\lambda/\mu),\quad
\bar{F}/\mu=\mathcal{O}(\lambda^2/\mu^2),\quad
\bar{G}/\mu=\mathcal{O}(\lambda/\mu),\quad
\bar{H}/\mu=\mathcal{O}(\lambda^3/\mu^3),
\end{equation}
and, specifically,
\be
(1-2\nu)\bar{E} \rightarrow \dfrac{4E}{3}, \qquad
(1-2\nu)\bar{G} \rightarrow \dfrac{E}{6}.
\en

For incompressible solids, $I_1=0$ for all stresses, and the fourth-order expansion Eq. \eqref{4thlog} reduces to
\be
W = \mu I_2 + \dfrac{\mathcal{A}}{3}I_3 + \mathcal{G}I_2^2,
\en
and only three constants remain\cite{Ogde74}.
Equivalently, in terms of the invariants of the Green strain tensor, the expansion of $W$ reads as Eq. \eqref{fourth-incomp}, where $\bar{D}$ is defined by Eqs. \eqref{set} or, equivalently, by
\be
\bar{D} = \mathcal{G} - \dfrac{\mathcal{A}}{2} + \dfrac{11\mu}{6},
\en
making it explicit that it is of the same order as $\mu$ (and thus, as $\bar{A}$):
\be
\bar{D}/\mu = \mathcal{O}(1).
\en

Turning our attention to the stress, we see that all the terms multiplying $\vec{\delta}$ in \eqref{stress4} are absorbed by the arbitrary hydrostatic stress, to give, in the limit,
\begin{equation}
\vec{t}=-p\vec{\delta}+2\mu\vec{e}+\mathcal{A}\vec{e}^2+4\mathcal{G}I_2\vec{e}.
\end{equation}
By the Cayley-Hamilton theorem we have (for an incompressible material, where $I_1=0$)
\begin{equation}
\vec{e}^3=\tfrac{1}{2}I_2\vec{e}+\tfrac{1}{3}I_3\vec{\delta},
\end{equation}
and hence we have
\begin{equation}
\vec{t}=-p\vec{\delta}+2\mu\vec{e}+\mathcal{A}\vec{e}^2+4\mathcal{G}(2\vec{e}^3-\tfrac{2}{3}I_3\vec{\delta}),
\end{equation}
and finally, by adjusting the hydrostatic term by introducing $p'=p+8\mathcal{G}I_3/3$,
\begin{equation}
\vec{t}=-p'\vec{\delta}+2\mu\vec{e}+\mathcal{A}\vec{e}^2+8\mathcal{G}\vec{e}^3,
\end{equation}
where only the three constants of fourth-order incompressible elasticity appear, and no invariant.
We recall that the corresponding measure of stress conjugate to the Green strain is given by the connection $\vec{\bar{t}}=(\vec{\delta}+2\vec{\bar{e}})^{-1}\vec{t}$, which applies for an isotropic material, yielding
\begin{equation}  \label{4stress-strain}
\vec{\bar{t}} = -p' (\vec{\delta}+2\vec{\bar{e}})^{-1} + 2\mu\vec{\bar{e}} + \bar{A} \vec{\bar{e}}^2 + 8 \bar{D}\vec{\bar{e}}^3.
\end{equation}
Note that the Lagrange multiplier $p'$ \emph{must} figure in the expressions for the stress (see also Eq.  \eqref{3stress-strain}), but 
has been omitted in the expression for the stress in several papers\cite{HaIZ04B, JCGB07, Woch08, Coul10}.

The counterpart of the strain--stress relation \eqref{e3} for the fourth order is very lengthy and is not written here.  We note, however, that the contribution to $\vec{e}$ additional to the first and second-order terms in \eqref{e3} for the third order in the stress has the structure
\begin{equation}
a_1\vec{t}^3+a_2T_1\vec{t}^2+(a_3T_1^2+a_4T_2)\vec{t}+(a_5T_3+a_6T_1T_2+a_7T_1^3)\vec{\delta},
\end{equation}
where $a_1,a_2,\dots ,a_7$ are constants that are collectively functions of $\mu$, $\lambda$ (or $E$, $\nu$), $\mathcal{A}$, $\mathcal{B}$, \dots, $\mathcal{H}$.
With the exception of $a_1=\mathcal{A}^2/16\mu^5$ these expressions are very lengthy and therefore omitted.
However, in the incompressible limit the strain-stress relation simplifies substantially and the extension of \eqref{e3incomp} becomes
\begin{multline}
\vec{e}=\frac{3}{4E^3}(2E^2+3\mathcal{A}T_1)(\vec{t}-\tfrac{1}{3}T_1\vec{\delta})
\\
-\frac{27\mathcal{A}}{8E^3}(\vec{t}^2-\tfrac{1}{3}T_2\vec{\delta})
+\frac{27}{32E^5}(T_1^2-3T_2)(8E\mathcal{G}-\mathcal{A}^2)(\vec{t}-\tfrac{1}{3}T_1\vec{\delta}).\label{e4incomp}
\end{multline}
As in the third-order case it is seen immediately that $\mbox{tr}(\vec{e})=0$.  In deriving \eqref{e4incomp} we have used the Cayley-Hamilton theorem for $\vec{t}$ to eliminate $\vec{t}^3$.


\section{EXAMPLES}
\label{examples}



\subsection{Acoustoelasticity of bulk acoustic waves}


Hughes and Kelly\cite{HuKe53} used the acousto-elastic effect to evaluate experimentally the third-order elasticity constants, by measuring the speed of infinitesimal bulk homogeneous plane waves propagating in a solid subject to a small pre-stress.
We summarize their results in Table I, where we use the layout of Norris\cite{Norr98} and the definition 
\be \label{K}
K = \dfrac{\text{d}(\rho v^2)}{\text{d}\sigma}\Big|_{\sigma=0}
\en
of the \emph{acousto-elastic coefficient} $K$,
where $\rho$ is the mass density in the unstrained state, and $\sigma$ is the pre-stress (which is either hydrostatic or uniaxial).

The incompressible counterparts to these formulas are readily established by use of Eqs. \eqref{lame-limit}  and \eqref{incomp_3rd}.
They appear in the last column of the table.
In particular, it is seen that the speed of longitudinal waves is infinite, as expected in an incompressible solid.
The formulas for shear waves are in accord with those established using different means by Gennisson \emph{et al.}\cite{Genn07} and by Destrade \emph{et al.}\cite{DeGS10} in the case of uniaxial pre-stress. For hydrostatic pre-stress, we note that the speeds of the shear waves are unaffected by the hydrostatic stress, which is also to be expected for an incompressible material.

\begin{sidewaystable}
 \begin{center}
\emph{Table I. Acousto-elastic coefficients for bulk acoustic waves in compressible and incompressible solids (here $3\kappa = 3\lambda + 2\mu$).}
\\[8pt]
  \begin{tabular}{ | l  | l | c | c |  c | c |}
    \hline \hline 
    stress  & mode   & propagation  & polarization & $K$ (compressible)  & $K$ (incompr.)
    \\ \hline \hline \rule[-20pt]{0pt}{46pt}
    hydrostatic   & longitudinal & arbitrary
    &  $ || \ \vec{n}$ &  $-\dfrac{1}{3\kappa}\left[7\lambda+10\mu+2\bar{A} + 10 \bar{B} + 6 \bar{C}\right]$ & $\infty$
    \\[12pt] \hline \rule[-20pt]{0pt}{46pt}
     hydrostatic   & transverse & arbitrary
    &  $ \perp \vec{n}$ &  $-\dfrac{1}{3\kappa}\left[3\lambda+6\mu+\bar{A} + 3 \bar{B} \right]$ & $0$
    \\[12pt] \hline \rule[-20pt]{0pt}{46pt}
   uni-axial   & longitudinal & $||$ stress
    &  $ || \ \vec{n}$ &  $-\dfrac{1}{3\kappa}\left[\lambda + 2\bar{B}  + 2\bar{C} +2 \dfrac{\lambda+\mu}{\mu}(2\lambda + 5\mu +\bar{A} + 2 \bar{B}) \right]$ & $\infty$
    \\[12pt] \hline \rule[-20pt]{0pt}{46pt}
  uni-axial   & longitudinal & $\perp$ stress
    &  $ || \ \vec{n}$ &  $-\dfrac{2}{3\kappa}\left[\bar{B}  + \bar{C} - \dfrac{\lambda}{\mu}(\lambda + 2\mu + \dfrac{\bar{A}}{2} +  \bar{B}) \right]$ & $\infty$
 \\[12pt] \hline \rule[-20pt]{0pt}{46pt}
  uni-axial   & transverse & $||$ stress
    &  $ \perp \vec{n}$ &  $-\dfrac{1}{3\kappa}\left[4(\lambda + \mu) +  \dfrac{\lambda+2\mu}{4\mu} \bar{A} +  \bar{B} \right]$ & $-\left[1+\dfrac{\bar{A}}{12\mu}\right]$
    \\[12pt] \hline \rule[-20pt]{0pt}{46pt}
      uni-axial   & transverse & $\perp$ stress
    &  $ ||$ stress  &  $-\dfrac{1}{3\kappa}\left[\lambda + 2\mu +  \dfrac{\lambda+2\mu}{4\mu} \bar{A} +  \bar{B} \right]$ & $ - \dfrac{\bar{A}}{12\mu}$
    \\[12pt] \hline \rule[-20pt]{0pt}{46pt}
      uni-axial   & transverse & $\perp$ stress
    &  $ \perp$ stress &  $\dfrac{1}{3\kappa}\left[2\lambda +  \dfrac{\lambda}{2\mu} \bar{A} -  \bar{B} \right]$ & $ 1+\dfrac{\bar{A}}{6\mu}$
    \\[12pt] \hline
  \end{tabular}
\end{center}
\end{sidewaystable}


\subsection{Acoustoelasticity of surface acoustic waves}


Hayes and Rivlin \cite{HaRi61} computed the acousto-elastic coefficient for surface acoustic wave (SAW) propagation.
For a wave propagating in the direction of uniaxial pre-stress of (small) magnitude $\sigma$, it is defined by Eq. \eqref{K} where $v$ is now the SAW speed.
Using the results of Tanuma and Man\cite{TaMa06, TaMa08}, we present it in the form 
\be
K = 1 - a\gamma_{22} - b\gamma_{23} - c\gamma_{33} - d\gamma_{44}
\en
for a compressible isotropic elastic solid,
where the $\gamma$'s are defined in terms of the Lam\'e constants by
\begin{align}
& \gamma_{22} = (\lambda + 2\mu)\left[-8(\lambda+\mu) + 2(5\lambda+6\mu)X - (2\lambda+3\mu)X^2\right]/\Delta, \notag \\
& \gamma_{23} = 4\lambda(1-X)\left[4(\lambda+\mu) - (\lambda+2\mu)X \right]/\Delta, \notag \\
& \gamma_{33} = \left[1-\mu X/(\lambda+2\mu)\right]/\Delta, \notag \\
& \gamma_{44} = -8(\lambda + 2\mu - \mu X)\left[2(\lambda+\mu) -(\lambda+2\mu)X\right]/\Delta, \notag \\
& \Delta = (\lambda + \mu)\left[8(3\lambda+4\mu) -16(\lambda+2\mu)X +3(\lambda+2\mu)X^2\right],
\end{align}
the non-dimensional quantity $X$ is the unique real positive root to Rayleigh's cubic\cite{Rayl85}
\be \label{rayleigh}
X^3 - 8X^2 + 8\dfrac{3\lambda+4\mu}{\lambda+2\mu}X - 16\dfrac{\lambda+\mu}{\lambda+2\mu}=0,
\en
and $a$, $b$, $c$, $d$ depend on the second- and third-order constants according to
\begin{align}
& a = \dfrac{(4\lambda + 3\mu)(\lambda + 2\mu)}{(3\lambda+2\mu)\mu} + \dfrac{2(\lambda + \mu)\bar{A}}{(3\lambda+2\mu)\mu} + \dfrac{2(2\lambda + 3\mu)\bar{B}}{(3\lambda+2\mu)\mu} + \dfrac{2\bar{C}}{3\lambda+2\mu},
\notag \\
& b = \dfrac{(\lambda + \mu)\lambda}{(3\lambda+2\mu)\mu} + \dfrac{(\lambda + 2\mu)\bar{B}}{(3\lambda+2\mu)\mu} + \dfrac{2\bar{C}}{3\lambda+2\mu},
\notag \\
& c = -\dfrac{(2\lambda + \mu)(\lambda + 2\mu)}{(3\lambda+2\mu)\mu} - \dfrac{\lambda\bar{A}}{(3\lambda+2\mu)\mu} - \dfrac{2(\lambda - \mu)\bar{B}}{(3\lambda+2\mu)\mu} + \dfrac{2\bar{C}}{3\lambda+2\mu},
\notag \\
& d = \dfrac{\lambda + \mu}{(3\lambda+2\mu)\mu} + \dfrac{(\lambda+2\mu) \bar{A}}{4(3\lambda+2\mu)\mu} + \dfrac{\bar{B}}{3\lambda+2\mu}.
\end{align}

Substantial simplifications occur when the incompressible limits  \eqref{lame-limit}--\eqref{incomp_3rd} apply. In particular, $X$ is now a definite number\cite{Rayl85}, namely $X= 0.9126$, the root of $X^3-8X^2+24X-16=0$, and $\gamma_{22} = -\gamma_{23}/2 = \gamma_{33} = 2(X-1)(4-X)/(26-16X+3X^2)$.
Because of these latter relationships, the limit of $K$ remains finite since, even though each of $a$, $b$, and $c$ goes to infinity as $\mathcal{O}(\lambda/\mu)$, the combination $a-2b+c$ tends to a finite limit, specifically $a-2b+c \rightarrow \bar{A}/(3\mu)$, whilst $d \rightarrow \bar{A}/(12\mu)$.
The final result is that the \emph{acousto-elastic coefficient for SAWs in incompressible media} is
\be
K = 1 + 0.9126 \dfrac{\bar{A}}{12 \mu},
\en
as established differently by Destrade \emph{et al.}\cite{DeGS10}.  We emphasize that this result applies for the situation in which the stress is uniaxial.  It is interesting to note in passing that in the corresponding plane strain problem (in the $(1,2)$ plane with stress $\sigma$ in the $x_1$ direction) it can be shown that $K=1-X/2$, which is independent of the third-order constant $\bar{A}$.


\subsection{Solitary waves in rods}


Porubov\cite{Poru03} showed that the propagation of nonlinear strain waves in an elastic rod with free lateral surface is governed by the so-called \emph{double-dispersive equation} (DDE).
For solids with the third-order strain energy density Eq. \eqref{third}, the DDE is
\be
v_{tt} - \alpha_1 v_{xx} - \alpha_2 \left(v^2\right)_{xx} - \alpha_3 v_{xxtt} + \alpha_4 v_{xxxx} = 0,
\en
for a strain function $v = v(x,t)$, where $x$ is the space variable in the direction of propagation, $t$ is time, subscripts denote partial differentiation, and
\be \label{poru3}
\alpha_1 = \dfrac{E}{\rho}, \qquad
\alpha_2 = \dfrac{\beta}{2\rho}, \qquad
\alpha_3 = \dfrac{\nu (\nu-1)}{2}R^2, \qquad
\alpha_4 = -\dfrac{\nu E R^2}{2\rho}.
\en
Here $E$ is Young's modulus, $\nu$ is Poisson's ratio, $R$ is the rod radius, $\rho$ is the mass density, and $\beta$ is the nonlinear parameter. (Note that this $\beta$ is different from the $\beta$ defined in Eqs. \eqref{alpha}.)
Explicitly,
\be
\beta = 3E + (1-2\nu)^3(\bar{B} + \bar{C}) + 2(1-2\nu)(1+\nu)(\bar{A} + 2\bar{B}) + 6 \nu^2 \bar{A}.
\en

Now, in the incompressible limit described by Eqs. \eqref{lame-limit}--\eqref{incomp_3rd}, this nonlinear parameter reduces to the very simple expression
\be
\beta = \tfrac{3}{2}(4 \mu + \bar{A}),
\en
and the $\alpha$'s simplify accordingly.
The analysis of Porubov can then be carried through, in particular to study solitary waves and solitons. It should be pointed out that the sign of $\beta$ can be determined for all incompressible third-order solids.
Indeed, it is well-known that, \emph{at the same level of approximation}, the model described by the Mooney--Rivlin strain energy density
\be
W = C_{10} [\text{tr} ( \vec{C} )- 3] + C_{01}[\text{tr} ( \vec{C}^{-1})-3],
\en
where $\vec{C} = 2\vec{\bar{e}} + \vec{\delta}$ is the right Cauchy-Green strain tensor and $C_{10}$ and $C_{01}$ are constants, is equivalent to the strain energy density Eq. \eqref{incomp_W}, with the connections
\be
\mu = 2(C_{10} + C_{01}), \qquad
\bar{A} = -8(C_{10} + 2 C_{01}).
\en
Clearly, it follows that here we have $\beta = -32 C_{01}$.
However, it is also well-known that the governing equations of motion for Mooney--Rivlin solids are strongly elliptic when\cite{Ogden1970} $C_{01}>0$.
Provided that this condition is satisfied, we deduce that
\be
\beta < 0,
\en
for all incompressible third-order solids and therefore that a compressive solitary wave emerges from an initial localized compressive input\cite{Poru03}.

In fact, Porubov and Maugin\cite{PoMa05} show that fourth-order elasticity is required to account for the possibility of simultaneous compressive and tensile solitary waves.
They use the Murnaghan\cite{Murn51} counterpart to Eq. \eqref{fourth}, where the expansion is carried out in terms of the principal invariants of $\vec{\bar e}$, which we write as
\be
i_1 = \text{tr} \ \vec{\bar e}, \qquad
i_2 = \tfrac{1}{2}\left[(\text{tr} \ \vec{\bar e})^2 - \text{tr} (\vec{\bar e}^2)\right], \qquad
i_3 = \text{det} \ \vec{\bar e}.
\en 
The Murnaghan strain-energy function is then given by
\be
W =
\dfrac{\lambda + 2\mu}{2} i_1^2 - 2\mu   i_2 + \dfrac{l+2m}{3} i_1^3 - 2m i_1 i_2 +  n i_3
+ \nu_1 i_1^4
+ \nu_2 i_1^2 i_2
+ \nu_3i_1i_3
+ \nu_4i_2^2,
\en
where $m$, $l$, \ldots, $\nu_4$ are constants.
The correspondence between the Landau and the Murnaghan constants is easy to establish. 
We find the connections
\begin{align} \label{murnLan}
& m = \dfrac{\bar{A}}{2} + \bar{B}, \qquad n = \bar{A}, \qquad l = \bar{B} + \bar{C}, \notag \\
& \nu_1 = \bar{E} + \bar{F} + \bar{G} + \bar{H}, \qquad
  \nu_2 = -3 \bar{E} - 2\bar{F} - 4\bar{G}, \qquad
  \nu_3 = 3\bar{E}, \qquad
  \nu_4 = 4 \bar{G},
\end{align}
and thus, the incompressible limits
\begin{align} \label{murnLim1}
& m/\mu = \mathcal{O}(\lambda/\mu), \qquad  m/\mu = \mathcal{O}(1), \qquad l/\mu = \mathcal{O}(\lambda^2/\mu^2), \notag \\
& \nu_1/\mu = \mathcal{O}(\lambda^3/\mu^3), \qquad
  \nu_2/\mu = \mathcal{O}(\lambda^2/\mu^2), \notag \\
  &
  \nu_3/\mu = \mathcal{O}(\lambda/\mu), \qquad
  \nu_4/\mu = \mathcal{O}(\lambda/\mu),
\end{align}
with the specific limits
\be \label{murnLim2}
(1-2\nu)m \rightarrow -E/3, \qquad
(1-2\nu)\nu_3 \rightarrow 4E, \qquad
(1-2\nu)\nu_4 \rightarrow 2E/3.
\en
Now, when fourth-order terms are taken into account, the corresponding DDE has an extra term\cite{PoMa05} and becomes
\be
v_{tt} - \alpha_1 v_{xx} - \alpha_2 \left(v^2\right)_{xx} - \alpha_3 v_{xxtt} + \alpha_4 v_{xxxx} - \alpha_5 (v^3)_{xx} = 0,
\en
where $\alpha_1$,  $\alpha_2$,  $\alpha_3$,  $\alpha_4$ are still given by Eqs. \eqref{poru3}, and $\alpha_5 = \gamma/(3\rho)$, with $\gamma$ (different from the $\gamma$ in Section  \ref{section3}) given by
\begin{multline}
E \gamma = E^2 - 8l^2(1-2\nu)^5(1+\nu) - 32m^2\nu^2(1-2\nu)(1+\nu)^3
\\ 
- 8n^2\nu^2(1-2\nu)(1+\nu) 
+ 4l(1-2\nu)^3\{E-4\nu(1+\nu)[2m(1+\nu)-n]\} \\
+ 8m(E+4n\nu^2)(1-2\nu)(1+\nu)^2  + 12nE\nu^2  
+ 8 \nu_1 E (1-2\nu)^4 \\ - 8\nu_2E(1-2\nu)^2(2-\nu)\nu + 8\nu_3E(1-2\nu)\nu^2 + 8\nu_4E(2-\nu)^2\nu^2.
\end{multline}
Clearly, the limits Eqs. \eqref{murnLim1}--\eqref{murnLim2} do not give a definite incompressible limit for $\gamma$.
In particular, the terms proportional to $(1-2\nu)m^2$ and to $\nu_4$ tend to infinity, and the limits of the terms proportional to  $(1-2\nu)^3 l m$ and to $(1-2\nu)^2\nu_2$ are not known. 
When $\gamma$ is written in terms of the Landau constants (using the inverse of Eq. \eqref{murnLan}), similar ambiguities arise.
However, if the expression for $\gamma$ is further transformed in terms of $\mathcal{A}$, $\mathcal{B}$, \ldots, $\mathcal{H}$, the nonlinear constants associated with the logarithmic strain, then it is a simple matter to find its unequivocal incompressible limit. 
In the end we find that $\gamma$ tends to 
\be
\gamma = \dfrac{11E}{3} -3\mathcal{A} + 18 \mathcal{G},
\en
or equivalently, in terms of the incompressible Landau constant $\bar{A}$ and the constant $\bar{D}$ defined in \eqref{set},
\be
\gamma = 14 \mu + 6\bar{A} + 18 \bar{D}.
\en


\subsection{Nonlinear plane waves}
\label{Nonlinear-plane-waves}


Wochner \emph{et al.}\cite{Woch08, WoHI08} also use the notation $\beta$ to identify the coefficient of cubic nonlinearity for \emph{shear waves}.  Again, this $\beta$ is different from that used previously in this paper, and is given by
\be
\beta = \dfrac{3}{4\mu}\left[\lambda + 2\mu + \bar A + 2 \bar B + 2 \bar G
-  \dfrac{\left(\lambda + 2\mu + \bar A/2 + \bar B \right)^2}{\lambda+\mu}\right].
\en
According to the first, second, and fifth limits in Eqs. \eqref{set}, this quantity should collapse to $\beta = 3(2\mu+ \bar A)/(4\mu)$ for incompressible solids.
In fact, the true limit is\cite{Woch08}
\be \label{beta-shear}
\beta = \dfrac{3}{2}\left( 1 + \dfrac{\bar A + 2\bar D}{2\mu}\right),
\en
because $\bar D$, as defined in Eq. \eqref{set}, should remain finite, although until now, this behavior had not been proved rigorously.
An alternative means of finding the correct limit is to rewrite the expression for $\beta$ in terms of the constants associated with the logarithmic strain, as
\be
\beta = \dfrac{3(1+\nu)}{2E}\left[ 2\mathcal{G} - \dfrac{E}{6(1+\nu)} - \dfrac{2(1+\nu)(1-2\nu)}{E} \left(\dfrac{\mathcal{A}}{2} + \mathcal{B} - \dfrac{E}{2(1+\nu)} \right)^2\right].
\en
Then the limit is clearly $\beta = 9(2\mathcal{G} - E/9)/(4E)$, which is the same as Eq. \eqref{beta-shear}.

Note that Wochner \emph{et al.}\cite{Woch08} also provide $\beta_l$, the Gol'dberg\cite{Gold61} coefficient of nonlinearity for \emph{longitudinal waves}, as
\be
\beta_l = \dfrac{3}{2} - \dfrac{\bar A + 3\bar B + \bar C}{\lambda + 2\mu}.
\en
If, as is incorrectly reported in Refs.\cite{HaIZ04, HaIZ07, OSIR07}, $\bar C$ were to behave as $\bar C/\mu = \mathcal{O}(\lambda/\mu)$, then $\beta_l$ would remain finite in the incompressible limit, suggesting that longitudinal homogeneous plane waves are possible.
As we have established in \eqref{incomp_3rd},  $\bar C/\mu$ behaves as $\mathcal{O}(\lambda^2/\mu^2)$, and $\beta_l$ therefore blows up in the incompressible limit, as should be expected, thereby precluding the existence of such waves.


\section{CONCLUDING REMARKS}


Using the logarithmic strain measure, we are able to determine the exact behavior of the elastic constants of second, third, and fourth orders in the incompressible limit, as collected below.
For the second-order Lam\'e constants
\be
\lambda \rightarrow \infty, \qquad \mu \rightarrow E/3,
\en
as is well known;
for third-order Landau constants,
\be
\bar A / \mu = \mathcal{O}(1), \qquad \bar B/\mu = \mathcal{O}(\lambda/\mu), \qquad \bar C /\mu = \mathcal{O}(\lambda^2/\mu^2),
\en
and for the fourth-order Landau constants,
\begin{align}
& \bar{E}/\mu=\mathcal{O}(\lambda/\mu),\qquad
\bar{F}/\mu=\mathcal{O}(\lambda^2/\mu^2),\notag \\
& \bar{G}/\mu=\mathcal{O}(\lambda/\mu),\qquad
\bar{H}/\mu=\mathcal{O}(\lambda^3/\mu^3).
\end{align}
For the constants which vary linearly with $\lambda$ as it goes to infinity, the specific limits are
\be
(1-2\nu)\bar{B} \rightarrow -E/3, \qquad
(1-2\nu)\bar{E} \rightarrow 4E/3, \qquad
(1-2\nu)\bar{G} \rightarrow E/6,
\en
as the Poisson ratio $\nu \rightarrow 1/2$.

We have used these limits to show that it is easy to take known results of elastic wave propagation in compressible materials to the corresponding incompressible limits.
In fact, other types of internal constraints could be accounted for just as easily\cite{DeSc04, EdFu09}.

We conclude the paper with two remarks.
The first is technical: it should not be forgotten that the hydrostatic term in the stress-logarithmic strain relation of an incompressible solid is an \emph{arbitrary} Lagrange multiplier, to be determined from initial/boundary conditions. 
This Lagrange multiplier also appears in the stress-Green strain relation, see Eqs. \eqref{3stress-strain} and \eqref{4stress-strain}.
Omission of this term would therefore lead to incorrect solutions of the equations.
The second is semantic: one gets the impression from reading the acoustics literature that `soft' and `incompressible' are two interchangeable adjectives.
It should be clear that they are not, and Nature and Engineering provide many examples of hard materials which are incompressible (such as fully saturated soils in undrained conditions\cite{DaSe96}) and of soft materials which are compressible (such as polyurethane foams\cite{BlKo62}).


\section*{Acknowledgments}

This work is supported by a Senior Marie Curie Fellowship awarded by the Seventh Framework Programme of the European Commission to the first author and by an E.T.S. Walton Award, given to the second author by Science Foundation Ireland.  This material is based upon works supported by the Science Foundation Ireland under Grant No. SFI 08/W.1/B2580.


\end{document}